\documentclass[
    10pt,
    a4paper,
    aps,
    longbibliography,
    nofootinbib,
    notitlepage,
    prb,
    superscriptaddress,
    twocolumn
]{revtex4-2}

\usepackage{amsmath}
\usepackage{amssymb}
\usepackage{amsthm}
\usepackage{csquotes}
\usepackage{microtype}
\usepackage{nicefrac}
\usepackage{orcidlink}
\usepackage{pifont}
\usepackage{physics}
\usepackage{siunitx}
\usepackage{textcomp}
\usepackage{txfonts}

\usepackage{graphicx}
\usepackage{float}
\usepackage{xcolor}
\usepackage{color, colortbl}
\usepackage{placeins}

\usepackage{capt-of} 
\usepackage{booktabs}
\usepackage{tabularx}
\usepackage{multirow}

\usepackage{hyperref}
\hypersetup{colorlinks=true, citecolor=blue, urlcolor=blue, linkcolor=blue}

\newcommand{\adj}[1]{#1^\dagger} 
\newcommand{\conj}[1]{#1^*} 
\newcommand{\trp}[1]{#1^\intercal} 

\newcommand{\vect}[1]{\boldsymbol{#1}} 
\newcommand{\uvect}[1]{\hat{\vect{#1}}} 
\newcommand{\matr}[1]{\boldsymbol{#1}} 
\newcommand{\adjvect}[1]{\adj{\vect{#1}}} 
\newcommand{\adjmatr}[1]{\adj{\matr{#1}}} 
\newcommand{\trpmatr}[1]{\trp{\matr{#1}}} 
\newcommand{\conjmatr}[1]{\conj{\matr{#1}}} 

\newcommand{\e}{\mathrm{e}} 
\newcommand{\mathpi}{\piup} 
\newcommand{\iu}{\mathrm{i}} 
\newcommand{\kb}{k_{\mathrm{B}}} 
\newcommand{\nbands}{N} 
\newcommand{\metricel}{G} 
\newcommand{\metric}{\matr{\metricel}} 
\newcommand{\hmatr}{\matr{H}} 
\newcommand{\ematr}{\matr{\mathcal{E}}} 
\newcommand{\bz}{\mathrm{BZ}} 
\newcommand{\spence}{\mathrm{Li}_2} 

\newcommand{\true}{\textcolor{green!80!black}{\ding{51}}} 
\newcommand{\false}{\textcolor{red}{\ding{55}}} 

\newcommand{\amend}[1]{#1}
\newenvironment{amendment}{}{}

\DeclareMathOperator{\diag}{diag}

\definecolor{Kitaev}{rgb}{1,0.45,0.46}
\definecolor{DMI}{rgb}{0.68,0.85,0.9}

\begin{document}
\title{Sign Changes in Heat, Spin, and Orbital Magnon Transport Coefficients in Kitaev Ferromagnets}

\author{Yannick Höpfner\orcidlink{0009-0004-9219-307X}}
\affiliation{Institut für Physik, Martin-Luther-Universität Halle-Wittenberg, D-06099 Halle (Saale), Germany}
\author{Ingrid Mertig\orcidlink{0000-0001-8488-0997}}
\affiliation{Institut für Physik, Martin-Luther-Universität Halle-Wittenberg, D-06099 Halle (Saale), Germany}
\author{Robin R.~Neumann\orcidlink{0000-0002-9711-3479}}
\email[Contact author: ]{rneumann@uni-mainz.de}
\affiliation{Institut für Physik, Johannes Gutenberg-Universität, D-55128 Mainz, Germany}
\affiliation{Institut für Physik, Martin-Luther-Universität Halle-Wittenberg, D-06099 Halle (Saale), Germany}

\begin{abstract}
Both Kitaev and Dzyaloshinskii-Moriya interactions (DMI) are known to promote intrinsic contributions to the magnon Hall effects such as the thermal Hall and the spin Nernst effects in collinear magnets.
Previously, it was reported that a sign change in those transversal transport coefficients only appears in the presence of Kitaev interaction, but not for DMI, which qualitatively distinguishes both kinds of spin-anisotropic interactions in ferromagnets.
Herein, we systematically study how the magnon-mediated heat, spin, and orbital transport in longitudinal and transverse geometries evolves with a continuously varying Kitaev-to-DMI ratio, but a fixed magnon band structure.
We show that several transport coefficients feature temperature-driven sign changes in the presence of Kitaev interaction, which are absent for DMI.
In particular, we find a sign change in \emph{longitudinal} orbital transport, the magnon orbital Seebeck effect, which is absent in the transverse geometry, the magnon orbital Nernst effect.
This sets the orbital transport apart from the heat and spin transport, where we only find sign changes promoted by the Kitaev interaction in \emph{transverse}, but not in the longitudinal geometry.

\end{abstract}
\date{\today}
\maketitle

\section{Introduction}

The Kitaev model, a compass-type Hamiltonian for $\nicefrac{1}{2}$ spins arranged on the honeycomb lattice, has become a paradigmatic example of a quantum spin liquid~%
\cite{
    kitaev_anyons_2006,
    nussinov_compass_2015%
}.
It has attracted much attention since it hosts anyons that are conceived to be relevant for (topological) quantum computing~%
\cite{
    kitaev_fault-tolerant_2003,
    lahtinen_short_2017%
}.
Among the candidate materials that realize the Kitaev interaction are the iridates Na$_2$IrO$_3$ and Li$_2$IrO$_3$~%
\cite{
    jackeli_mott_2009,
    chaloupka_kitaev-heisenberg_2010,
    choi_spin_2012,
    singh_relevance_2012,
    chaloupka_zigzag_2013,
    rau_generic_2014,
    hwan_chun_direct_2015%
},
the cobaltates Na$_2$Co$_2$TeO$_6$ and Na$_3$Co$_2$SbO$_6$~%
\cite{
    liu_pseudospin_2018,
    sano_kitaev-heisenberg_2018,
    liu_kitaev_2020,
    songvilay_kitaev_2020%
},
the transition-metal halide $\alpha$-RuCl$_3$~%
\cite{
    plumb_alpha-rucl_3_2014,
    kim_kitaev_2015,
    banerjee_proximate_2016,
    ran_spin-wave_2017%
},
and the van-der-Waals chromium trihalide CrI$_3$~%
\cite{
    xu_interplay_2018,
    chen_topological_2018,
    soriano_magnetic_2020,
    stavropoulos_magnetic_2021,
    chen_magnetic_2021%
}.
Because these materials also exhibit Heisenberg interactions, the magnetic ground state remains ordered down to very low temperatures, giving rise to collective spin excitations known as magnons.
The Kitaev interaction is imprinted on the magnon band structure and their wave functions.
Conventionally, inelastic neutron scattering has been employed to experimentally determine the spin interactions including the Kitaev interaction.
However, it has been noticed that Dzyaloshinskii-Moriya interaction (DMI)~%
\cite{%
    dzyaloshinsky_thermodynamic_1958,
    moriya_anisotropic_1960%
}
appears naturally in these honeycomb lattices and can give rise to similar features in the band structure complicating the quantification of the Kitaev interaction.

\begin{figure}
    \centering
    \includegraphics[width=\linewidth]{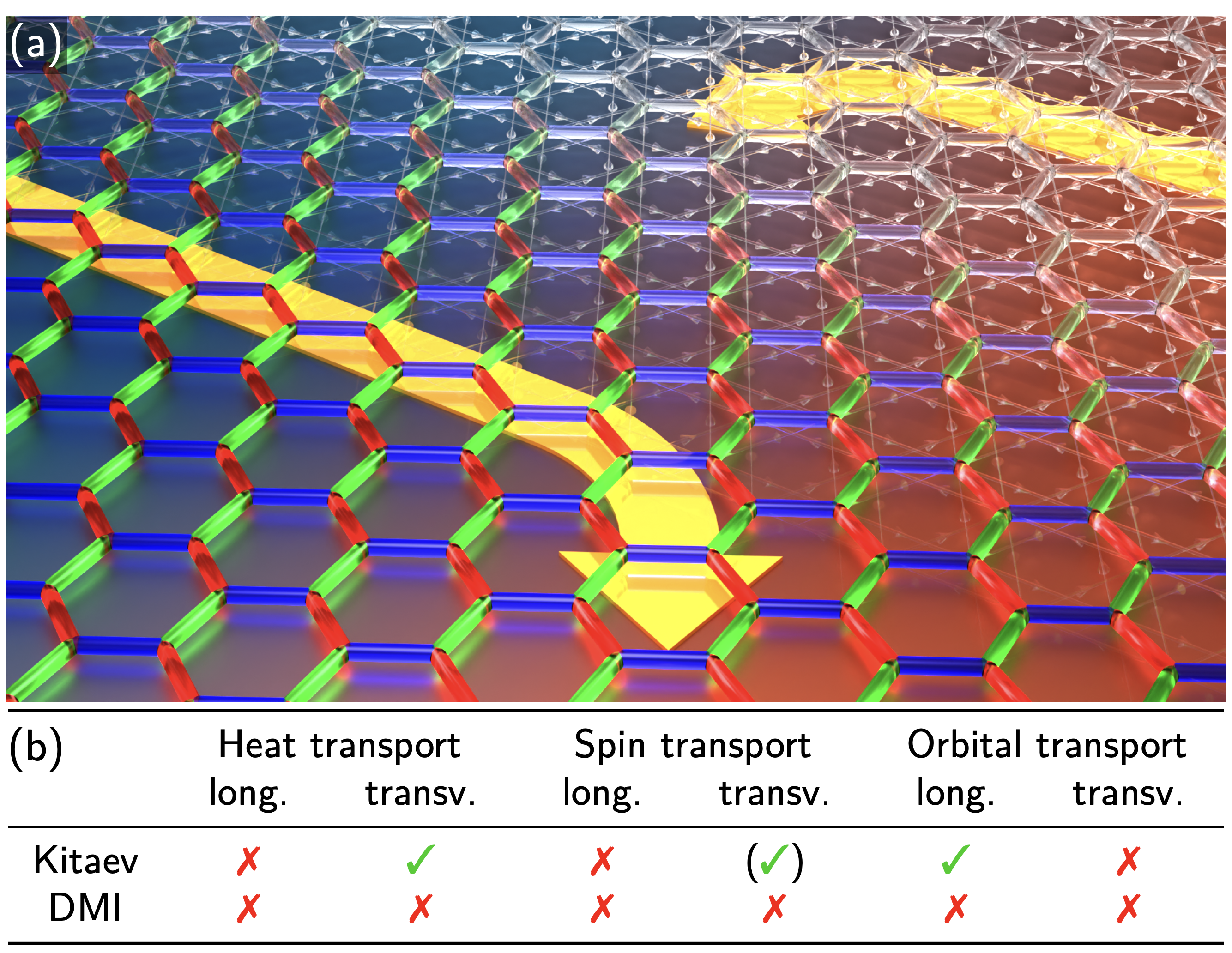}
    \caption{
        (a) Orbital moment currents of magnons (yellow arrows) induced by a temperature gradient in honeycomb ferromagnets hosting Kitaev (red/blue/green-colored hexagons) and Dzyaloshinskii-Moriya interactions (transparent hexagons).
        For the selected mean temperature, the longitudinal currents are antiparallel, while the transverse currents are parallel in the two cases.
        (b) Table of heat, spin, and orbital transport coefficients in longitudinal (long.) and transverse (transv.) geometries.
        Check marks (\true) and cross marks (\false) indicate the presence and absence of a temperature-driven sign change in the corresponding transport coefficient, respectively.
        For the transversal spin transport, the sign change is strongly suppressed.
    }
    \label{fig:graphical_abstract}
\end{figure}
For ferromagnets, the thermal Hall effect (THE) and the spin Nernst effect (SNE) were suggested as probes to distinguish both interactions based on the presence or absence of a temperature-driven sign change~\cite{zhang_interplay_2021}.
In this work, we compare the consequences of the DMI and the Kitaev interaction on the heat, spin, and, orbital transport in longitudinal and transverse geometries.
As an example, we consider the ferromagnet CrI$_3$, which we describe by a series of parameter combinations involving varying ratios of DMI and Kitaev interaction by fitting its experimentally obtained magnon band structure.
Considering the transport of the magnon orbital moment, we predict that the longitudinal (transverse) currents may be antiparallel (parallel) in the cases of Kitaev interaction and DMI [cf.~Fig.~\ref{fig:graphical_abstract}(a)].
In general, we find that no temperature-driven sign changes appear with DMI, while the Kitaev interaction promotes sign changes for the transverse heat [thermal Hall effect (THE)] and spin transport [spin Nernst effect (SNE)], as well as for the longitudinal orbital transport [orbital Seebeck effect (OSE)].
In contrast, the longitudinal heat (Fourier's law) and spin transport [spin Seebeck effect (SSE)] and the transverse orbital transport [orbital Nernst effect (ONE)] do not change sign.
These findings are summarized in Fig.~\ref{fig:graphical_abstract}(b).

We trace back the sign change in the THE to the Berry curvature, which qualitatively changes in the presence of Kitaev interaction by developing a low-energy contribution of opposite sign close to $\Gamma$.
By decomposing the Berry curvature into its contributions originating from the other bands, we demonstrate the vital role of the virtual \enquote{hole-like} bands that result from the breaking of magnon number conservation due to the Kitaev interaction.

A previous study has shown a behavior of the SNE very similar to the THE~\cite{zhang_interplay_2021}.
However, below we contrast those findings with our results indicating a strong suppression of the sign change in the SNE.
We explain the difference by contrasting the Berry curvature with the \emph{spin Berry curvature}, which we compute to account for the breaking of spin conservation.

While the sign change of the spin Berry curvature is not pronounced but exists, it is completely absent in the orbital Berry curvature, which undergoes only minor corrections if DMI is substituted by Kitaev interaction.
Notwithstanding, the magnon orbital moment texture in reciprocal space features both signs for the Kitaev interaction and thereby distinguishes between DMI and Kitaev interaction.
This texture is shown to be responsible for the observed sign change of the OSE.

\section{Results}

\begin{figure*}
    \centering
    \includegraphics[width=\linewidth]{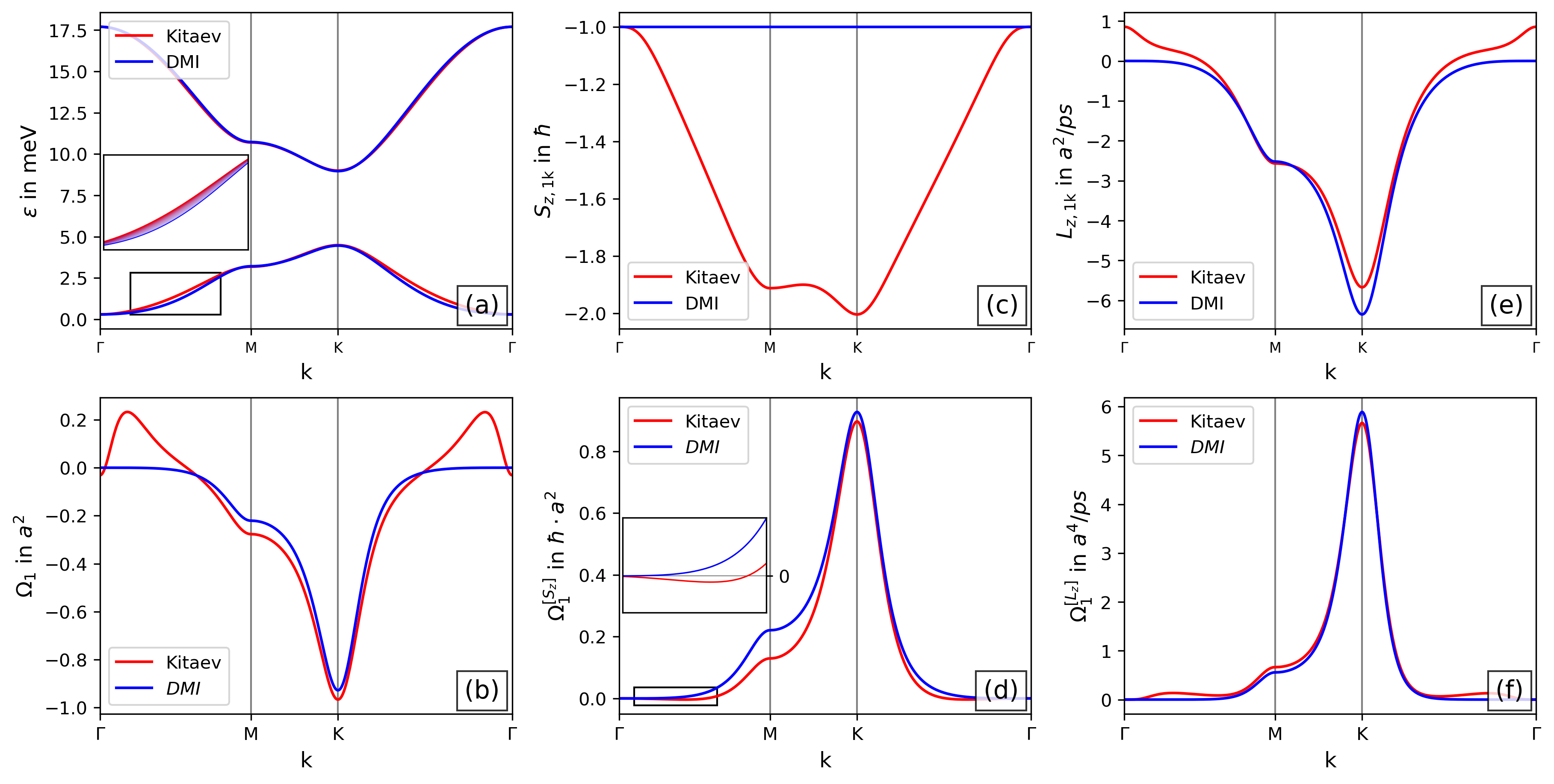}
    \caption{
    (a) The magnon band structure, (b) the Berry curvature, (c) the spin, (d) the spin Berry curvature, (e) the orbital moment, and (f) the orbital Berry curvature along a high-symmetry path  in $\vect{k}$ space for the DMI (blue curves) and Kitaev models (red curves).
    In panels (b--f) only the lowest band is depicted.
    In both models, the Chern numbers are $+1$ and $-1$ for the lower and the upper band, respectively.
    The parameters for the DMI model read
    $
        J_1 = \SI{-2.076}{\milli\electronvolt},
        J_2 = \SI{0.169}{\milli\electronvolt},
        J_3 = \SI{0.143}{\milli\electronvolt},
        A = \SI{-0.1}{\milli\electronvolt},
        D = \SI{0.289}{\milli\electronvolt},
        K = \SI{0}{\milli\electronvolt}
    $,
    and those of the Kitaev model are
    $
        J_1 = \SI{-0.2}{\milli\electronvolt},
        J_2 = \SI{0}{\milli\electronvolt},
        J_3 = \SI{0}{\milli\electronvolt},
        A = \SI{-0.1}{\milli\electronvolt},
        D = \SI{0}{\milli\electronvolt},
        K = \SI{-5.2}{\milli\electronvolt}
    $.
    Here, $a$ is the lattice constant.
    }
    \label{fig:microscopics}
\end{figure*}

\begin{figure*}
    \centering
    \includegraphics[width=\linewidth]{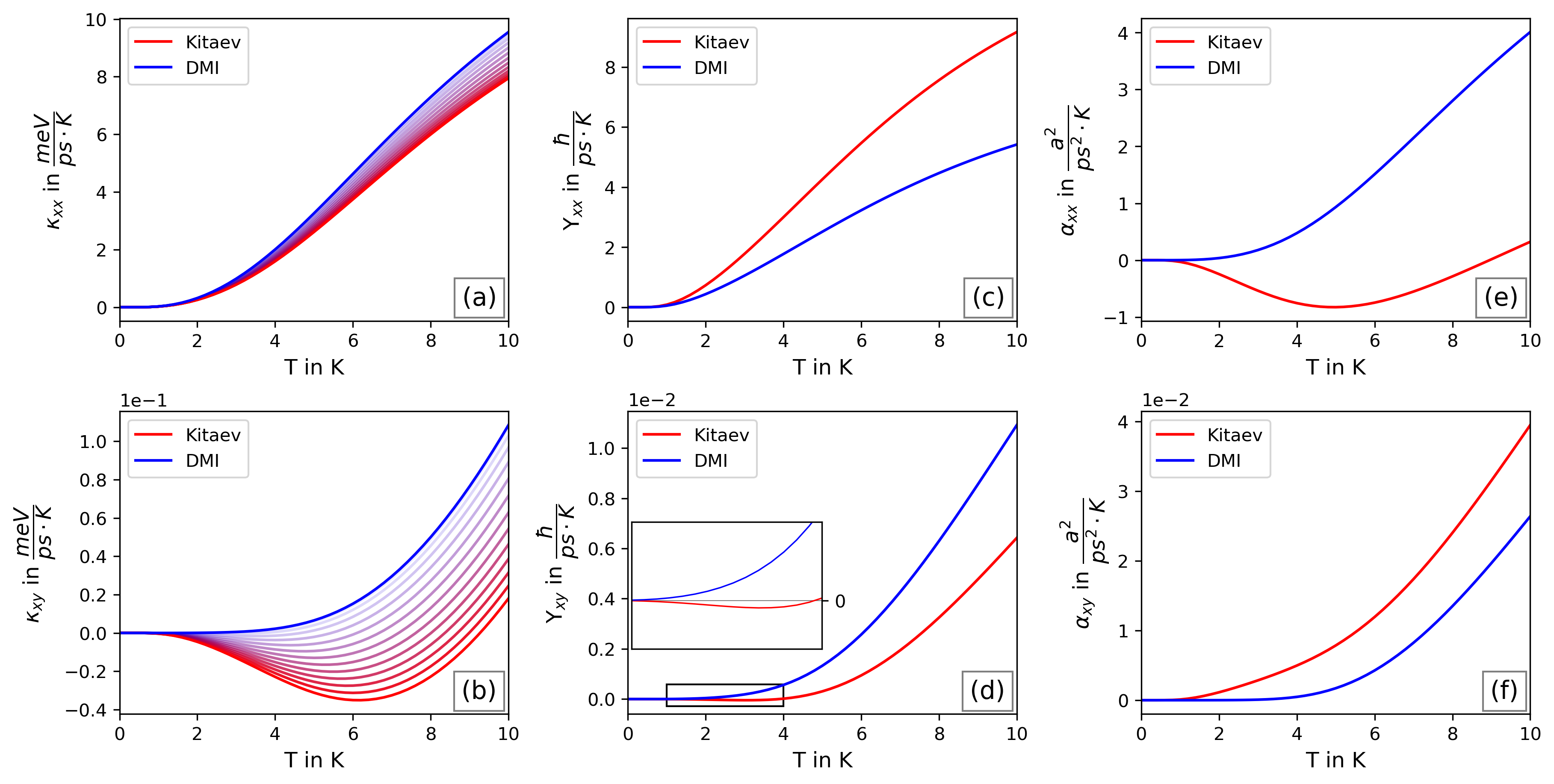}
    \caption{
        Heat, spin, and orbital transport coefficients versus temperature.
        (a,~b) The longitudinal and transverse heat conductivities for different parameter sets from Table~\ref{tab:params}.
        (c,~d) The spin Seebeck and Nernst coefficients.
        (e,~f) The orbital magnon Seebeck and Nernst coefficients.
        The blue and red curves refer to the DMI and Kitaev models, respectively.
        For the relaxation time, we assume $\tau = \SI{100}{\pico\second}$.
    }
    \label{fig:transport}
\end{figure*}

We consider a monolayer of a \amend{stacked} honeycomb ferromagnet as realized in, e.g., van der Waals magnets.
Typically, the symmetry of these materials admits both DMI and Kitaev interactions~%
\cite{%
    moriya_anisotropic_1960,
    cheng_spin_2016,
    mcclarty_topological_2018%
}.
Thus, we study the Hamiltonian
\begin{align}
    \begin{split}
    H
    &=
    \sum_{r=1}^3
    \frac{J_{r}}{2 \hbar^2}
    \sum_{\langle i j \rangle_r}
    \vect{S}_i \cdot \vect{S}_j
    +
    \frac{1}{2 \hbar^2}
    \sum_{\langle i j \rangle_2}
    \vect{D}_{ij} \cdot
    \left(\vect{S}_i \times \vect{S}_j \right)
    \\
    &\quad
    +
    \frac{K}{2 \hbar^2}
    \sum_{\langle ij \rangle_1}
    \qty(\vect{S}_i \cdot \uvect{\gamma}_{ij})
    \qty(\vect{S}_j \cdot \uvect{\gamma}_{ij})
    +
    \frac{A}{\hbar^2}
    \sum_i \qty(S_i^z)^2
    .
    \end{split}
    \label{eq:shamil}
\end{align}
It comprises Heisenberg exchange up to \amend{third-nearest} neighbors
$
    \langle ij \rangle_r
$
($r = 1, 2, 3$),
DMI with
$
    \vect{D}_{ij} = \pm D \uvect{z}
$,
where the $+$ ($-$) sign holds for counterclockwise (clockwise) bond orientation [cf.~Fig.~\ref{fig:graphical_abstract}(a)], pointing perpendicular to the lattice.
The Kitaev interaction $K$ is defined with respect to the orthogonal axes
$
    \uvect{\gamma}_{ij}
$,
whose directions are locked to the respective bond directions.%
\footnote{%
    In the monolayer of CrI$_3$, the vectors $\uvect{\gamma}_{ij}$ are defined as the normal vectors of the corresponding Cr$_2$I$_2$ plaquette~%
    \cite{%
        lee_fundamental_2020,
        aguilera_topological_2020,
        zhang_interplay_2021%
    }.
}
Furthermore, the easy-axis anisotropy $A < 0$ stabilizes the spins along the out-of-plane direction in their ferromagnetic ground state.
Note that we omit $\varGamma$ and $\varGamma'$ interactions that generally accompany the Kitaev interaction~%
\cite{%
    katukuri_kitaev_2014,
    rau_generic_2014,
    chaloupka_hidden_2015%
},
whose effect on the magnon band structure can alternatively be obtained by renormalizing the remaining parameters \amend{for out-of-plane polarized magnets}~\cite{mcclarty_topological_2018}.

The Hamiltonian is recast using the truncated Holstein-Primakoff transformation~\cite{holstein_field_1940}
\begin{align}
    \frac{S_i^+}{\hbar}
    =
    \sqrt{2 S} a_i
    ,
    \quad
    \frac{S_i^-}{\hbar}
    =
    \sqrt{2 S} \adj{a}_i
    ,
    \quad
    \frac{S_i^z}{\hbar}
    =
    S - \adj{a}_i a_i
    ,
\end{align}
which, after Fourier transformation, results in a bilinear bosonic Bogoliubov-de Gennes Hamiltonian
\begin{align}
    H
    &=
    \frac{1}{2}
    \sum_{\vect{k}}
    \adjvect{\phi}_{\vect{k}}
    \hmatr_{\vect{k}}
    \vect{\phi}_{\vect{k}}
\end{align}
with
$
    \adjvect{\phi}_{\vect{k}}
    =
    \mqty(
        \adj{a}_{1 \vect{k}} & \cdots & \adj{a}_{\nbands \vect{k}}
        &
        a_{1 (-\vect{k})} & \cdots & a_{\nbands (-\vect{k})}
    )
$
being a Nambu spinor and $\nbands$ being the number of bands.
Note that the Nambu-space description becomes necessary only in the presence of Kitaev interaction because the off-diagonal blocks of $\hmatr_{\vect{k}}$, which correspond to the anomalous pairing terms
$
    a_{m \vect{k}}
    a_{n, -\vect{k}}
$
and
$
    \adj{a}_{m \vect{k}}
    \adj{a}_{n, -\vect{k}}
$,
vanish for $K = 0$.
Thus, $K$ breaks the conservation of the particle number.
The analytic expression of $\hmatr_{\vect{k}}$ for the Hamiltonian in Eq.~\eqref{eq:shamil} is given in Appendix~\ref{sec:hamil}.
Then, a Bogoliubov transformation~%
\cite{%
    bogoljubov_new_1958,
    valatin_comments_1958,
    colpa_diagonalization_1978,
    shindou_topological_2013%
}
\begin{subequations}
\begin{align}
    \adjmatr{T}_{\vect{k}}
    \hmatr_{\vect{k}}
    \matr{T}_{\vect{k}}
    &=
    \ematr_{\vect{k}}
    =
    \diag\mqty(
        \varepsilon_{1 \vect{k}} & \cdots & \varepsilon_{\nbands \vect{k}}
        &
        \varepsilon_{1 (-\vect{k})} & \cdots & \varepsilon_{\nbands (-\vect{k})}
    )
    ,
    \label{eq:bogtrafo_energies}
    \\
    \adjmatr{T}_{\vect{k}}
    \metric
    \matr{T}_{\vect{k}}
    &=
    \metric
    =
    \diag\mqty(1 & \cdots & 1 & -1 & \cdots & -1)
    \label{eq:bogtrafo_paraunitary}
\end{align}
\end{subequations}
defines new eigenmodes as
$
    \adjvect{\psi}_{\vect{k}}
    =
    \adjvect{\phi}_{\vect{k}}
    \qty(\adjmatr{T}_{\vect{k}})^{-1}
    =
    \mqty(
        \adj{\alpha}_{1 \vect{k}} & \cdots & \adj{\alpha}_{\nbands \vect{k}}
        &
        \alpha_{1 (-\vect{k})} & \cdots & \alpha_{\nbands (-\vect{k})}
    )
$
and transforms the Hamiltonian into
$
    H
    =
    \sum_{\vect{k}}
    \sum_{n = 1}^{\nbands}
    \varepsilon_{n \vect{k}}
    \adj{\alpha}_{n \vect{k}} \alpha_{n \vect{k}}
$.
Constants have been omitted because they solely shift the ground-state energy.

Henceforth, the parameters of CrI$_3$ are chosen.
For all investigated parameter sets, we fix $S = \nicefrac{3}{2}$.
The material's magnon band structure has been described both using DMI and Kitaev interactions~%
\cite{%
    chen_topological_2018,
    lee_fundamental_2020,
    chen_magnetic_2020,
    chen_magnetic_2021,
    cen_determining_2023,
    kim_spin_2024%
}.
As a starting point, we use the parameters $J_1 = \SI{-0.2}{\milli\electronvolt}$, $K = \SI{-5.2}{\milli\electronvolt}$, $A = \SI{-0.1}{\milli\electronvolt}$, and $D = J_2 = J_3 = 0$~\cite{lee_fundamental_2020}.
As $D = 0$, we refer to this parameter set as Heisenberg-Kitaev or, in short, Kitaev model.
Its magnon band structure, shown in red in Fig.~\ref{fig:microscopics}(a), is not exclusive to this particular model.
We have identified a range of parameter sets combining DMI and Kitaev interaction, which approximately preserve the band structure.
They have been obtained by varying $K$ and fitting $J_1$, $J_2$, $J_3$, and $D$ such that the magnon energies at the high-symmetry points $\Gamma$, M, and K are retained.
($A$ was left unchanged because it is uniquely determined by the spin-wave gap.)
In particular, a parameter set involving no Kitaev interaction was determined as
$
    J_1 = \SI{-2.076}{\milli\electronvolt},
    J_2 = \SI{0.169}{\milli\electronvolt},
    J_3 = \SI{0.143}{\milli\electronvolt},
    A = \SI{-0.1}{\milli\electronvolt},
    D = \SI{0.289}{\milli\electronvolt},
    \text{ and }
    K = \SI{0}{\milli\electronvolt}
$.
This Heisenberg-DMI or, in short, DMI model reproduces the same band structure apart from minor deviations [blue curve in Fig.~\ref{fig:microscopics}(a)].
As depicted in the inset, the band structures of the mixed parameter sets lie between these limiting cases.
In Appendix~\ref{sec:param_path} we give the complete list of parameters and show how the parameters evolve with $K$.

These results demonstrate that it cannot be concluded if Kitaev interaction is actually present in the system and how large it is based on the magnon band structure, which is in agreement with previous studies~%
\cite{%
    zhang_interplay_2021,
    chen_magnetic_2021,
    brehm_topological_2024%
}.
Despite the lack of differences in the \emph{band structure} between DMI and Kitaev interaction, it has been pointed out that the DMI and the Kitaev interaction can be distinguished based on the thermal Hall effect~%
\cite{%
    mcclarty_topological_2018,
    aguilera_topological_2020,
    zhang_interplay_2021%
}.
In the following, we systematically study the evolution of the magnon transport coefficients along the path in parameter space while leaving the band structure unchanged.

In linear response theory, the heat, spin, and orbital currents are written as
\begin{subequations}
\begin{align} 
    \vect{j}_{\text{h}} &= \matr{\kappa} (-\grad T)
    ,
    \\
    \vect{j}_{\text{s}} &= \matr{\varUpsilon} (-\grad T)
    ,
    \\
    \vect{j}_{\text{o}} &= \matr{\alpha} (-\grad T)
    ,
\end{align}
\end{subequations}
where $\vect{j}_{\text{h}}$, $\vect{j}_{\text{s}}$, and $\vect{j}_{\text{o}}$ correspond to the heat, spin, and orbital current densities, respectively, which are driven by a temperature gradient $\grad T$.
For spin and orbital currents, the spin and orbital polarizations are taken along the $z$ axis.
The heat conductivity
\begin{align}
    \matr{\kappa}
    &=
    \begin{pmatrix}
        \kappa_{xx} & \kappa_{xy}
        \\
        -\kappa_{xy} & \kappa_{xx}
    \end{pmatrix}
\end{align}
only features two independent elements corresponding to Fourier's law ($\kappa_{xx}$) and the thermal Hall effect ($\kappa_{xy}$) due to the three-fold rotational symmetry, which also applies to the thermal spin conductivity $\matr{\varUpsilon}$ featuring the spin Seebeck ($\varUpsilon_{xx}$) and the spin Nernst effects ($\varUpsilon_{xy}$), and the thermal orbital conductivity $\matr{\alpha}$ featuring the orbital Seebeck ($\alpha_{xx}$) and the orbital Nernst effect ($\alpha_{xy}$)~\cite{seemann_symmetry-imposed_2015}.

First, we focus on the heat transport.
The two independent transport coefficients can be computed as~%
\cite{%
    callaway_quantum_1991,
    katsura_theory_2010,
    matsumoto_rotational_2011,
    kovalev_thermomagnonic_2012,
    matsumoto_thermal_2014,
    mook_taking_2018%
}
\begin{subequations}
\begin{align}
    \kappa_{xx}
    &=
    \frac{\tau}{V T}
    \sum_{n = 1}^{\nbands}
    \sum_{\vect{k}}
    \varepsilon_{n \vect{k}}^2
    v_{x, n \vect{k}}^2
    \eval{
        \qty(
            -\pdv{\rho}{\varepsilon}
        )
    }_{\varepsilon_{nk}}
    ,
    \label{eq:kxx}
    \\
    \kappa_{xy}
    &=
    - \frac{\kb^2 T}{V \hbar}
    \sum_{n = 1}^{\nbands}
    \sum_{\vect{k}}
    \varOmega_{n \vect{k}} c_{2}[\rho(\varepsilon_{nk})]
    ,
    \label{eq:kxy}
\end{align}
\end{subequations}
where $\hbar$ is the Planck constant, $\tau$ is the (constant) relaxation time,%
\footnote{%
    In the following, we estimate the phenomenological relaxation time for heat, spin, and orbital transport coefficients to be \SI{100}{\pico\second}.%
}
$V$ is the system's total volume/area,
$
    v_{x, n \vect{k}}
    =
    (1 / \hbar)
    \pdv*{\varepsilon_{n \vect{k}}}{k_x}
$
is the group velocity along $x$, $\rho$ is the Bose distribution, and
$
    c_2(\rho)
    =
    (1 + \rho) \ln^2 \qty(\frac{1 + \rho}{\rho})
    -
    \ln^2(\rho)
    -
    2 \spence(-\rho)
$
with the dilogarithm $\spence$.
$\kappa_{xy}$ involves the Berry curvature defined as~%
\cite{%
    berry_quantal_1984,
    matsumoto_thermal_2014,
    mook_thermal_2019%
}
\begin{subequations}
\begin{align}
    \varOmega_{n \vect{k}}
    &=
    \sum\limits_{\underset{n \neq m}{m = 1}}^{2 \nbands}
    \varOmega_{n m \vect{k}}
    ,
    \\
    \varOmega_{n m \vect{k}}
    &=
    \iu \hbar^2
    \sum_{\mu, \nu = x, y}
    \epsilon_{\mu \nu}
    \frac{
        \qty(\metric \matr{v}_{\mu, \vect{k}})_{nm}
        \qty(\metric \matr{v}_{\nu, \vect{k}})_{mn}
    }{
        \qty(
            \tilde{\varepsilon}_{n \vect{k}}
            -
            \tilde{\varepsilon}_{m \vect{k}}
        )^2
    }
    ,
\end{align}
\label{eq:berry_curv}
\end{subequations}
where $\epsilon_{\mu \nu}$ is the Levi-Civita symbol, 
$
    \matr{v}_{\lambda \vect{k}}
    =
    (1 / \hbar)
    \adjmatr{T}_{\vect{k}}
    \qty(
        \partial_{k_{\lambda}}
        \hmatr_{\vect{k}}
    )
    \matr{T}_{\vect{k}}
$
are the matrix elements of the group velocity operator in the eigenbasis of $H$, and
$
    \tilde{\varepsilon}_{l \vect{k}}
    =
    \qty(
        \metric
        \ematr_{\vect{k}}
    )_{ll}
$
are the signed magnon energies.

Because the expression for $\kappa_{xx}$ only contains properties that depend on the magnon \emph{band structure}, which are nearly identical [cf.~Fig.~\ref{fig:microscopics}(a)], we can expect $\kappa_{xx}$ to be similar for all parameter sets.
Indeed, as presented in Fig.~\ref{fig:transport}(a), $\kappa_{xx}$ increases monotonically with temperature and does not qualitatively differ between the parameter sets.
Quantitative differences solely originate from the imperfect agreement of the band structures.

In contrast, the expression for $\kappa_{xy}$ additionally contains the Berry curvature, which we plot in Fig.~\ref{fig:microscopics}(b).
We restrict ourselves to the lower band in the following because those states govern the transport at low temperatures.
In the DMI model, the Berry curvature is exclusively negative in the vicinity of the K points and vanishes around $\Gamma$.
In the Kitaev model, the Berry curvature additionally exhibits a positive contribution in the vicinity of $\Gamma$, which is absent in the DMI model.
Hence, while in the DMI model the Berry curvature possesses only one sign, it exhibits both signs in the Kitaev model.
Although the negative Berry curvature is larger and determines the Chern number
$
    C_1
    =
    -\frac{1}{2 \mathpi}
    \int_{\bz} \varOmega_{1 \vect{k}} \dd[2]{k}
$,
which are identical for the DMI and Kitaev models as they are adiabatically connected (i.e., the band gap does not close), the positive Berry curvature is located at lower energies, which may open up the possibility to probe it at low temperatures.

To understand the qualitative differences in the Berry curvatures $\varOmega_{n \vect{k}}$ between the DMI and the Kitaev models, we have decomposed it into its individual contributions $\varOmega_{n m \vect{k}}$ induced by the other bands $m$.
Note that the bands 3 and 4 are \enquote{virtual copies} of bands 1 and 2 that emerge due to the Bogoliubov-de Gennes formalism [cf.~Eq.~\eqref{eq:bogtrafo_energies}].
Focusing on the lower band ($n = 1$), in the DMI model the Berry curvature is only induced by the particle bands ($m \leq 2$), but has no contributions from hole bands ($m > 2$).
This is because the DMI model does not break the conservation of the magnon number rendering the Nambu-space description redundant.
In other words, the DMI model can exactly be described as a two-band model.
On the other hand, in the Kitaev model $\varOmega_{1 \vect{k}}$ features both negative contributions from particle and positive contributions from hole bands.
More details can be found in Appendix~\ref{sec:berry}.

The hole-band induced contributions to the Berry curvature are reflected in the thermal Hall effect [cf.~Fig.~\ref{fig:transport}(b)].
$\kappa_{xy}$ increases monotonically for the DMI model, but features a minimum and a sign change for the Kitaev model.
For intermediate models combining the DMI and the Kitaev interaction, the minimum is more pronounced the larger the Kitaev-to-DMI ratio.
As the minimum becomes deeper, it is shifted towards higher temperatures.
The $c_2$ function acts as a modified occupation function, due to which the positive Berry curvature states in the Kitaev model are favored at temperatures up to around \SI{9}{\kelvin}.
Above, the larger negative contributions stemming from higher-energy magnon states are no longer frozen out, thus resulting in a sign change of $\kappa_{xy}$.
Hence, above a certain activation temperature, the negative Berry curvature, which governs $C_1$, eventually dominates $\kappa_{xy}$.
This temperature-driven sign change sets the DMI and the Kitaev interaction apart.

Next, we analyze the spin transport.
Here, the transport coefficients are computed as
\cite{
    cheng_spin_2016,
    zyuzin_magnon_2016,
    mook_taking_2018,
    mook_spin_2019,
    li_intrinsic_2020%
}
\begin{subequations}
\begin{align}
    \varUpsilon_{xx}
    &=
    \frac{\tau}{2 V T}
    \sum_{n = 1}^{2 \nbands}
    \sum_{\vect{k}}
    \metricel_{nn}
    \qty(J_{x, \vect{k}}^{[S_z]})_{nn}
    \varepsilon_{n \vect{k}}
    v_{x, n \vect{k}}
    \eval{
        \qty(
            -\pdv{\rho}{\varepsilon}
        )
    }_{\metricel_{nn} \varepsilon_{nk}}
    ,
    \\
    \varUpsilon_{xy}
    &=
    \frac{\kb}{V \hbar}
    \sum_{n = 1}^{\nbands}
    \sum_{\vect{k}}
    \varOmega_{n \vect{k}}^{[S_z]}
    c_1[\rho(\varepsilon_{n\vect{k}})]
    ,
\end{align}
\end{subequations}
with
$
    c_1(x)
    =
    (1 + x) \ln (1 + x) - x \ln x
$.
We have introduced the $O$-current density operator, whose matrix elements 
$
    \matr{J}_{\beta, \vect{k}}^{[O]}
    =
    \qty(
        \matr{O}_{\vect{k}}
        \metric
        \matr{v}_{\beta, \vect{k}}
        +
        \matr{v}_{\beta, \vect{k}}
        \metric
        \matr{O}_{\vect{k}}
    ) / 2
$
($\beta = x, y$)
enter the generalized $O$-Berry curvature~%
\cite{%
    zyuzin_magnon_2016,
    li_intrinsic_2020%
}
\begin{align}
    \varOmega_{n \vect{k}}^{[O]}
    &=
    \amend{-\hbar^2
    \Im}
    \sum_{\mu, \nu = x, y}
    \epsilon_{\mu \nu}
    \sum_{\underset{m \neq n}{m = 1}}^{2 \nbands}
    \frac{
        \qty(
            \metric
            \matr{J}_{\mu, \vect{k}}^{[O]}
        )_{nm}
        \qty(
            \metric
            \matr{v}_{\nu, \vect{k}}
        )_{mn}
    }{
        \qty(
            \tilde{\varepsilon}_{n \vect{k}}
            -
            \tilde{\varepsilon}_{m \vect{k}}
        )^2
    }
    .
    \label{eq:generalized_berry_curv}
\end{align}
Note that contrary to the conventional Berry curvature [Eq.~\eqref{eq:berry_curv}] that enters the Chern number, the generalized Berry curvature is generally not associated with a topological invariant.

For the spin Seebeck ($\varUpsilon_{xx}$) and spin Nernst effects ($\varUpsilon_{xy}$), $O$ is substituted by $S_z$ whose matrix elements in the Hamiltonian's eigenbasis read
$
    \matr{O}_{\vect{k}}
    =
    \matr{S}_{z, \vect{k}}
    =
    -\hbar
    \adjmatr{T}_{\vect{k}}
    \matr{T}_{\vect{k}}
$~\cite{okuma_magnon_2017}.
In Fig.~\ref{fig:microscopics}(c), we present the spin expectation values
$
    S_{z, n \vect{k}}
    \coloneqq
    \qty(S_{z, \vect{k}})_{nn}
$
for the lower band.
While the $z$ component of the total spin operator commutes with the Hamiltonian in the DMI model~\cite{cheng_spin_2016}, it is not conserved in the presence of Kitaev interaction.
Consequently, the magnons have a quantized spin expectation value of
$
    S_{z, n \vect{k}}
    \equiv
    -\hbar
$
in the DMI model, but a non-quantized expectation value of up to $2 \hbar$ in the Kitaev model.
This also shows in the microscopic spin current, which is up to two times larger (cf.~Appendix~\ref{sec:currents}).
Therefore, one can expect a relatively higher efficiency for longitudinal spin transport in the Kitaev model.

In Fig.~\ref{fig:transport}(c) we have plotted $\varUpsilon_{xx}$ as a function of temperature, which increases monotonically.
$\varUpsilon_{xx}$ is larger for the Kitaev than for the DMI model and the difference is more pronounced than for $\kappa_{xx}$, which demonstrates that this difference cannot be exclusively explained by deviations in the band structures, but is related to the spin.

Another consequence of broken spin conservation is that the spin Berry curvature
$
    \varOmega_{n \vect{k}}^{[S_z]}
$
is not merely a product of the spin expectation value and the Berry curvature in the Kitaev model [cf.~Fig.~\ref{fig:microscopics}(d)].
Although this is correct for the DMI model, in the Kitaev model the sign change in $\varOmega_{1 \vect{k}}^{[S_z]}$ is strongly suppressed and one observes a deviation between both models at M.
If one would simply compute the product between spin expectation value and Berry curvature, one would obtain a \emph{larger} spin Berry curvature at M for the Kitaev model than for the DMI model, which is in contradiction to our findings.
Accordingly, the nondiagonal elements of the spin current operator strongly modify $\varOmega_{1 \vect{k}}^{[S_z]}$ for the Kitaev model.
The decomposition of the spin Berry curvature demonstrates that the difference at M originates from a positive hole band-induced contribution (cf.~Appendix~\ref{sec:berry}).
Hence, despite the larger spin expectation value of the magnon states in the Kitaev model, the spin Berry curvature is smaller.

As a result, $\varUpsilon_{xy}$ is smaller for the Kitaev model than for the DMI model [cf.~Fig.~\ref{fig:transport}(d)].
This observation could already be expected from the behavior of $\kappa_{xy}$.
However, in contrast to $\kappa_{xy}$, there is no pronounced sign change in $\varUpsilon_{xy}$ for the Kitaev model, which obstructs the possibility of qualitatively discerning both kinds of spin-anisotropic interactions.
Instead, within the two-current model, in which the spin Berry curvature is replaced by the product of spin expectation value and Berry curvature, the sign change is clearly resolvable~\cite{zhang_interplay_2021}.

The orbital Seebeck and orbital Nernst effects
\cite{%
    bhowal_orbital_2021,
    pezo_orbital_2022,
    busch_orbital_2023,
    go_magnon_2024%
}
\begin{subequations}
\begin{align}
    \alpha_{xx}
    &=
    \frac{\tau}{2 V T}
    \sum_{n = 1}^{2 \nbands}
    \sum_{\vect{k}}
    \metricel_{nn}
    \qty(J_{x, \vect{k}}^{[L_z]})_{nn}
    \varepsilon_{n \vect{k}}
    v_{x, n \vect{k}}
    \eval{
        \qty(
            -\pdv{\rho}{\varepsilon}
        )
    }_{\metricel_{nn} \varepsilon_{nk}}
    ,
    \\
    \alpha_{xy}
    &=
    \frac{\kb}{\hbar V}
    \sum_{n = 1}^{\nbands}
    \sum_{\vect{k}}
    c_1[\rho(\varepsilon_{n \vect{k}})]
    \varOmega_{n \vect{k}}^{[L_z]}
\end{align}
\end{subequations}
are calculated in close analogy to $\varUpsilon_{xx}$ and $\varUpsilon_{xy}$ apart from the spin operator, which is substituted by the matrix elements
\begin{align}
    \begin{split}
    \qty(L_{z, \vect{k}})_{n m}
    &=
    \sum_{\mu, \nu = x, y}
    \frac{\iu \hbar \epsilon_{\mu \nu}}{2}
    \sum_{\underset{n \neq l \neq m}{l = 1}}^{2 \nbands}
    \qty(
        \frac{1}{
            \tilde{\varepsilon}_{l \vect{k}}
            -
            \tilde{\varepsilon}_{m \vect{k}}
        }
        +
        \frac{1}{
            \tilde{\varepsilon}_{l \vect{k}}
            -
            \tilde{\varepsilon}_{n \vect{k}}
        }
    )
    \\
    &\quad
    \times
    \metricel_{nn}
    \metricel_{ll}
    \qty(v_{\mu, \vect{k}})_{nl}
    \qty(v_{\nu, \vect{k}})_{lm}
    ,
    \end{split}
    \label{eq:orbital_moment}
\end{align}
of the orbital moment operator
$
    \vect{L}
    =
    \qty(
        \vect{r} \times \vect{v}
        -
        \vect{v} \times \vect{r}
    ) / 2
$~%
\cite{%
    pezo_orbital_2022,
    busch_orbital_2023,
    go_magnon_2024,
    an_intrinsic_2025%
}.%
\footnote{%
    \label{ftn:orbital_moment}
    Note that our definition of the orbital moment operator of magnons deviates from Refs.~\cite{go_magnon_2024,an_intrinsic_2025} by an additional bosonic metric $G_{nn}$.
    It ensures that the operator $L_z$ is particle-hole symmetric, i.e.,
    $
        \matr{L}_{z, \vect{k}}
        =
        \matr{\varSigma}_{x}
        \trpmatr{L}_{z, -\vect{k}}
        \matr{\varSigma}_{x}
    $,
    where
    $
        \matr{\varSigma}_{x}
        =
        \matr{\sigma}_x
        \otimes
        \matr{I}
    $
    is the $2 \nbands \times 2 \nbands$ matrix that interchanges particle and hole sectors.
    Here, $\matr{\sigma}_x$ is the $2 \times 2$ Pauli matrix and $\matr{I}$ is the $\nbands \times \nbands$ identity matrix.
    The particle-hole symmetry is enforced to restore the analogy to the spin operator, which is also particle-hole symmetric, and adopt its linear-response expressions.%
}

In Fig.~\ref{fig:microscopics}(e) we show the orbital moment
$
    L_{z, n \vect{k}}
    \coloneqq
    \qty(L_{z, \vect{k}})_{nn}
$
of the lowest band along a high-symmetry path in the first Brillouin zone.
In case of the DMI model, $L_{z, 1 \vect{k}}$ vanishes at $\Gamma$, has a saddle point at M, an extremum at K and only features the negative sign.
For the Kitaev model, it changes to positive values at and close to $\Gamma$.
This sign change does not carry over to the orbital Berry curvature, shown for the lowest band in Fig.~\ref{fig:microscopics}(f).
Here, $\varOmega_{1 \vect{k}}^{[L_z]}$ exclusively assumes positive values for both models and there is no trace of the sign change in $L_{z, 1 \vect{k}}$.
This qualitatively distinguishes the orbital from the conventional and the spin Berry curvatures.

Turning to the transport coefficients, the sign change in $L_{z, 1 \vect{k}}$ manifests in a sign change in $\alpha_{xx}$ for the Kitaev model, while it is monotonic for the DMI model.
This is because the orbital moment enters $\alpha_{xx}$ via the orbital current, which can be qualitatively distinct from the group velocity (cf.~Appendix~\ref{sec:currents}).

On the other hand, $\alpha_{xy}$ is governed by the orbital Berry curvature and, hence, does not exhibit the sign change as seen in Fig.~\ref{fig:transport}(f); it increases monotonically with temperature for both the DMI and Kitaev models.
Although the DMI curve is suppressed for a larger temperature range, a qualitative feature distinguishing both curves is missing.

Our findings on the orbital Berry curvature are in part in agreement with the work by An and Kim~\cite{an_intrinsic_2025}.
They also find low-energy peaks in $\varOmega_{1 \vect{k}}^{[L_z]}$ in the presence of Kitaev interaction, however, their sign is anisotropic and depends on the direction of $\vect{k}$.
Moreover, their orbital Berry curvature breaks the three-fold rotational symmetry, while it is preserved in our calculations.
Furthermore, they find that $\alpha_{xy}$ may have a different sign in the Kitaev and the DMI models.
These differences are potentially caused by (i) an additional metric in Eq.~\eqref{eq:orbital_moment} and (ii) and the definition of the generalized Berry curvature in Eq.~\eqref{eq:generalized_berry_curv}, where we compute the full cross product (expressed by the Levi-Civita symbol) of the generalized current and the group velocity.
\begin{amendment}
    Modification (i) restores the particle-hole symmetry of $\vect{L}$ so that the linear response expressions derived for spin transport can be applied~(cf. footnote~\ref{ftn:orbital_moment}).
    Modification (ii) is implemented to extract the \emph{antisymmetric} component of the transport tensors, e.g.,
    $
        (\alpha_{xy} - \alpha_{yx}) / 2
    $,
    which is associated with the Hall effect, while the symmetric component, e.g.,
    $
        (\alpha_{xy} + \alpha_{yx}) / 2
    $,
    is related to an anisotropic longitudinal conductivity and corresponds to a pseudo-Hall effect~%
    \cite{%
        goldberg_new_1954,
        koch_notizen_1955%
    }.
\end{amendment}

\section{Discussion}
Our calculations demonstrate that the magnon thermal Hall effect ($\kappa_{xy}$), the orbital Seebeck effect ($\alpha_{xx}$), and, at least in principle, the spin Nernst effect ($\varUpsilon_{xy}$) allow one to distinguish DMI and Kitaev interaction based on the presence or absence of a temperature-induced sign change, while Fourier's law ($\kappa_{xx}$), the spin Seebeck effect ($\varUpsilon_{xx}$), and the orbital Nernst effect ($\alpha_{xy}$) do not provide clear distinguishing features.
\begin{amendment}
    In principle, these results suggest that, for a given band structure, one can quantify the magnitudes of the DMI and the Kitaev interaction based on transport properties.
    For example, one could relate the temperature $T_0$ at which the sign change of $\kappa_{xy}$ takes place with the magnitude of the Kitaev interaction.
    Generally, $T_0$ increases with the Kitaev interaction.
    In practice, however, the exact details depend on the band structure and magnons may not be the only heat carriers contributing to $\kappa_{xy}$.
    Furthermore, other mechanisms such as skew scattering or side jump due to impurities or many-body interactions exist~%
    \cite{%
        mangeolle_thermal_2022,
        mangeolle_phonon_2022,
        oh_phonon_2025,
        chatzichrysafis_thermal_2025%
    }
    and their dependence on the Kitaev-to-DMI ratio is unclear.
    These limitations hamper the precise quantification of the DMI and the Kitaev interaction.
\end{amendment}

\begin{amendment}
    Throughout this work, we have considered the particular band structure of CrI$_3$ as an example and have neglected $\varGamma$ and $\varGamma'$ interactions, which can be expected to renormalize the transport properties~%
    \cite{%
        cookmeyer_spin-wave_2018,
        chern_sign_2021,
        zhang_topological_2021,
        koyama_field-angle_2022,
        li_magnons_2023,
        chern_topological_2024,
        mosadeq_unveiling_2024%
    }.
    Reference~\cite{mosadeq_unveiling_2024} reports that the sign change may be suppressed those interactions, although the authors study the out-of-plane field-polarized phase of $\alpha$-RuCl$_3$, which is a zigzag antiferromagnet in its ground state without a magnetic field~\cite{sears_magnetic_2015}.
    Still, $\kappa_{xy}$ exhibits a non-monotonic behavior with a negative minimum, which may be the remainder of the sign change and is clearly different from the results that we have obtained for the DMI.
    Additional calculations are necessary to assess the robustness of the sign change of the corresponding magnon transport coefficients induced by the Kitaev interaction under various conditions.

    Speculatively, one can expect that some qualitative features of the Kitaev interaction survive as it brings about the anomalous pairing terms that lift the magnon number conservation.
\end{amendment}
The resulting Bogoliubov-de Gennes Hamiltonian allows for additional contributions to the conventional, spin, and orbital Berry curvatures as well as the orbital moment.
As we have explicitly shown for the Berry curvature, these additional contributions can be traced back to the virtual magnon bands and may give rise to sign changes \emph{within one band} for these $\vect{k}$-dependent quantities.
These contributions fundamentally set magnons apart from electrons, since the electron number is conserved.

The absence of a clear distinguishing feature in the orbital Nernst effect for the DMI and Kitaev interaction could intuitively be explained by the common wisdom that the orbital Hall and orbital Nernst effects are known to exist even without spin-orbit coupling~%
\cite{%
    bernevig_orbitronics_2005,
    go_intrinsic_2018,
    jo_gigantic_2018,
    bhowal_orbital_2021,
    pezo_orbital_2022,
    go_magnon_2024%
}.
However, this does \emph{not} imply that these two forms of spin-orbit coupling, DMI and Kitaev interaction, are irrelevant for magnon orbitronics.
In fact, our calculations have revealed a complex orbital texture featuring a sign change with Kitaev interaction that is absent with DMI.
Apart from the orbital Seebeck effect, one could envision that this sign change in $L_{z, 1 \vect{k}}$ could be also uncovered as a macroscopic net orbital moment in thermal equilibrium~\cite{neumann_orbital_2020} or in the orbital Edelstein effect of magnons~\cite{li_magnonic_2020}, thereby revealing its nontrivial texture for the Kitaev model.

However, since magnons possess neither a charge nor a mass, the \enquote{orbital moment of magnons} neither entails an orbital magnetic moment nor an orbital angular momentum.
Thus, it remains an open question of how to probe the orbital moment of magnons.
It has been suggested to employ the electric polarization or the electrical voltage due to the orbital motion of magnons as probe~%
\cite{%
    matsumoto_theoretical_2011,
    matsumoto_rotational_2011,
    go_magnon_2024,
    neumann_electrical_2024,
    to_theory_2024%
},
however, this has the shortcoming that it relies on relativistic magnetoelectric coupling~%
\cite{%
    katsura_spin_2005,
    tokura_multiferroics_2014%
}
and that it also includes contributions from spin currents unrelated to the magnon orbital motion~\cite{meier_magnetization_2003}.
It remains unclear whether the magnon orbital moment couples to electronic orbital angular momentum or chiral phonons.
These are questions that should be addressed in future studies.

\section*{Data availability}
The data that support the findings of this article are openly available~\cite{zenodo}.

\begin{acknowledgments}
This work was funded by the Deutsche Forschungsgemeinschaft (DFG, German Research Foundation) -- Project-ID 328545488 -- TRR 227, project B04.
\end{acknowledgments} 


\appendix
\section{Magnon Hamiltonian}
\label{sec:hamil}
The matrix $\hmatr_{\vect{k}}$ can generally be written as
\begin{align}
    \hmatr_{\vect{k}}
    &=
    \begin{pmatrix}
        \matr{A}_{\vect{k}} & \matr{B}_{\vect{k}}
        \\
        \conjmatr{B}_{-\vect{k}} & \conjmatr{A}_{-\vect{k}}
    \end{pmatrix}
    ,
\end{align}
where the relations
$
    \matr{A}_{\vect{k}} = \adjmatr{A}_{\vect{k}}
$
and
$
    \adjmatr{B}_{\vect{k}} = \conjmatr{B}_{-\vect{k}}
$
ensure the Hermiticity of the Hamiltonian.
For the spin Hamiltonian in Eq.~\eqref{eq:shamil}, it can be written as
\begin{align}
    \matr{A}_{\vect{k}}
    &=
    S
    \begin{pmatrix}
            a_{\vect{k}}
            -
            \iu D d_{\vect{k}}
        &
            (J_1 + K/3) f_{1 \vect{k}}
            +
            J_3 f_{3 \vect{k}}
        \\
            (J_1 + K/3) \conj{f}_{1 \vect{k}}
            +
            J_3 \conj{f}_{3 \vect{k}}
        &
            a_{\vect{k}}
            +
            \iu D d_{\vect{k}}
    \end{pmatrix}
    ,
    \\
    \matr{B}_{\vect{k}}
    &=
    S
    \begin{pmatrix}
        0
        &
            K b_{\vect{k}} / 3
        \\
            K b_{-\vect{k}} / 3
        &
        0
    \end{pmatrix}
    ,
\end{align}
where
\begin{align}
    a_{\vect{k}}
    &=
    -3 J_1
    +
    J_2 (f_{2 \vect{k}} - 6)
    -
    3 J_3
    -
    K
    +
    2 A
    \\
    f_{i \vect{k}}
    &=
    \sum_{\vect{\delta}_i}
    \e^{\iu \vect{k} \cdot \vect{\delta}_i}
    ,
    \\
    d_{\vect{k}}
    &=
    \sum_{\vect{\delta}_2}
    \nu(\vect{\delta}_2)
    \e^{\iu \vect{k} \cdot \vect{\delta}_2}
    ,
    \\
    b_{\vect{k}}
    &=
    \sum_{\vect{\delta}_1}
    \e^{\iu [\vect{k} \cdot \vect{\delta}_1 - \phi(\vect{\delta}_1)]}
    .
\end{align}
Here, the sum over $\vect{\delta}_i$ runs over all $i$-th nearest neighbors of a spin on sublattice 1,
$
    \nu(\vect{\delta}_2)
    =
    \pm 1
$
depending on whether the bond vector $\vect{\delta}_2$ is oriented along the counterclockwise ($+$) or clockwise ($-$) sense of rotation, and
$
    \phi(\vect{\delta}_1)
$
is the azimuthal angle of the bond vector $\vect{\delta}_1$, i.e.,
$
    \uvect{\delta}_1
    =
    \trp{\mqty(
        \cos \phi(\vect{\delta}_1)
        &
        \sin \phi(\vect{\delta}_1)
        &
        0
    )}
$.

\section{Fitted parameter sets}
\label{sec:param_path}

\begin{figure}
    \centering
    \includegraphics[width=\linewidth]{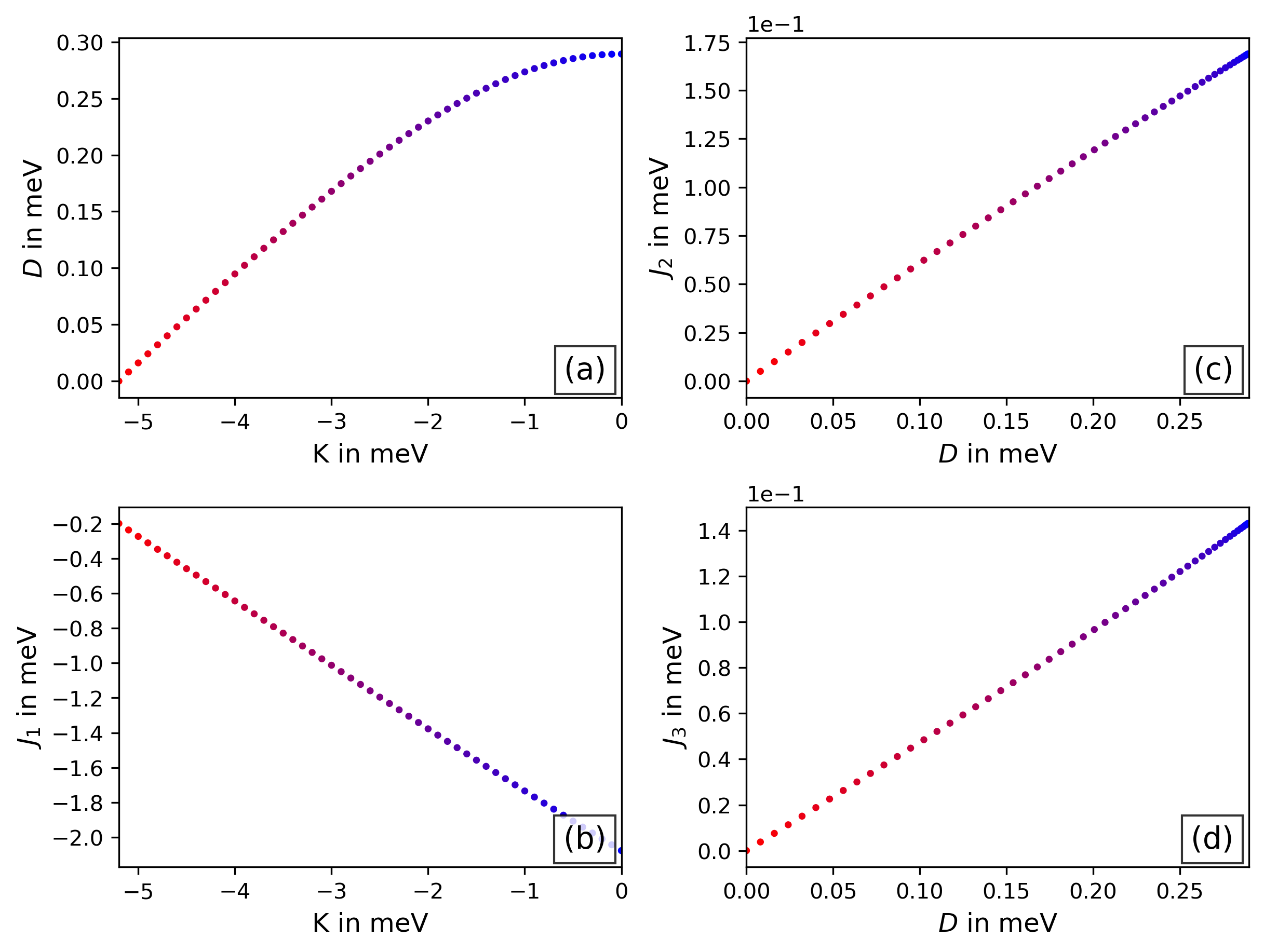}
    \caption{
        Evolution of the parameters of the spin Hamiltonian [cf.~Eq.~\eqref{eq:shamil}] (a) $D$, (b) $J_1$, (c) $J_2$, and (d) $J_3$ with $K$.
        The parameters are determined to (approximately) yield the same magnon band structures as in Fig.~\ref{fig:microscopics}(a).
    }
    \label{fig:param_path}
\end{figure}
Taking the band structure of the Kitaev model as a starting point, we have fitted the magnon energies at the high-symmetry points $\Gamma$, K, and M by ramping down $\abs{K}$ from its initial value (\SI{5.2}{\milli\electronvolt}) to zero.
Since only $A$ determines the spin-wave gap for the Hamiltonian in Eq.~\eqref{eq:shamil}, it was fixed at its initial value.
Therefore, five energies (one at $\Gamma$, two at K and M) have been fitted with four parameters ($J_1$, $J_2$, $J_3$, and $D$).
\amend{The results are listed in Table~\ref{tab:params}.}
Although the fitting procedure was restricted to the high-symmetry points, the magnon band structures of all parameter sets approximately coincide in the entire Brillouin zone.
The largest deviation is observed between the DMI and the Kitaev models.
Between the K and the $\Gamma$ point, where the deviations are the most prominent, the energy difference is below \SI{0.3}{\milli\electronvolt} between all parameter sets [cf.~inset in Fig.~\ref{fig:microscopics}(a)].

The interdependence of the parameters is visualized in Fig.~\ref{fig:param_path}.
As $\abs{K}$ is varied from \SIrange{5.2}{0}{\milli\electronvolt}, $D$ increases from \SIrange{0}{0.3}{\milli\electronvolt}, i.e., $D$ substitutes $K$ [Fig.~\ref{fig:param_path}(a)].
Although both interactions open a Haldane gap between the two magnon bands at K, $K$ additionally increases the total band width.
In order to fix the total band width, $\abs{J_1}$ needs to increase as $\abs{K}$ decreases [Fig.~\ref{fig:param_path}(b)].
Furthermore, the asymmetry between the upper and lower bands caused by $K$ is neither achieved by $D$ nor $J_1$.
Hence, $J_2$ and $J_3$ have to fulfill this role as $K$ is replaced by $D$ [Fig.~\ref{fig:param_path}(e,~f)].

\begin{widetext}
\begin{minipage}{\textwidth}
    \centering
    \captionof{table}{
        Parameter sets for the Hamiltonian in Eq.~\eqref{eq:shamil} that approximately yield the same magnon band structure.
        The highlighted parameter sets belong to the Kitaev (red) and the DMI model (blue).
    }
    \begin{minipage}{.5\linewidth}
        \centering
        \begin{tabular}{|c|c|c|c|c|c|}
            \hline
            \multicolumn{1}{|c|}{$K$} &
            \multicolumn{1}{c|}{$D$} &
            \multicolumn{1}{c|}{$A$} &
            \multicolumn{1}{c|}{$J_1$} &
            \multicolumn{1}{c|}{$J_2$} &
            \multicolumn{1}{c|}{$J_3$} \\
            \hline
            \rowcolor{Kitaev}-5.2 & 0.000 & -0.1 & -0.200 & 0.000 & 0.000 \\
            -5.1 & 0.008 & -0.1 & -0.237 & 0.005 & 0.004 \\
            -5.0 & 0.016 & -0.1 & -0.274 & 0.010 & 0.008 \\
            -4.9 & 0.024 & -0.1 & -0.311 & 0.015 & 0.011 \\
            -4.8 & 0.032 & -0.1 & -0.348 & 0.020 & 0.015 \\
            -4.7 & 0.040 & -0.1 & -0.385 & 0.025 & 0.019 \\
            -4.6 & 0.048 & -0.1 & -0.422 & 0.030 & 0.023 \\
            -4.5 & 0.056 & -0.1 & -0.459 & 0.034 & 0.026 \\
            -4.4 & 0.064 & -0.1 & -0.496 & 0.039 & 0.030 \\
            -4.3 & 0.072 & -0.1 & -0.533 & 0.044 & 0.034 \\
            -4.2 & 0.079 & -0.1 & -0.570 & 0.049 & 0.037 \\
            -4.1 & 0.087 & -0.1 & -0.607 & 0.053 & 0.041 \\
            -4.0 & 0.095 & -0.1 & -0.644 & 0.058 & 0.045 \\
            -3.9 & 0.102 & -0.1 & -0.681 & 0.062 & 0.048 \\
            -3.8 & 0.110 & -0.1 & -0.718 & 0.067 & 0.052 \\
            -3.7 & 0.117 & -0.1 & -0.755 & 0.071 & 0.056 \\
            -3.6 & 0.125 & -0.1 & -0.792 & 0.076 & 0.059 \\
            -3.5 & 0.132 & -0.1 & -0.829 & 0.080 & 0.063 \\
            -3.4 & 0.140 & -0.1 & -0.866 & 0.084 & 0.066 \\
            -3.3 & 0.147 & -0.1 & -0.903 & 0.088 & 0.070 \\
            -3.2 & 0.154 & -0.1 & -0.939 & 0.093 & 0.073 \\
            -3.1 & 0.161 & -0.1 & -0.976 & 0.097 & 0.077 \\
            -3.0 & 0.168 & -0.1 & -1.013 & 0.101 & 0.080 \\
            -2.9 & 0.175 & -0.1 & -1.050 & 0.104 & 0.084 \\
            -2.8 & 0.181 & -0.1 & -1.086 & 0.108 & 0.087 \\
            -2.7 & 0.188 & -0.1 & -1.123 & 0.112 & 0.090 \\
            -2.6 & 0.194 & -0.1 & -1.159 & 0.116 & 0.093 \\
            \hline
        \end{tabular}
    \end{minipage}%
    \begin{minipage}{.5\linewidth}
        \centering
        \begin{tabular}{|c|c|c|c|c|c|}
            \hline
            \multicolumn{1}{|c|}{$K$} &
            \multicolumn{1}{c|}{$D$} &
            \multicolumn{1}{c|}{$A$} &
            \multicolumn{1}{c|}{$J_1$} &
            \multicolumn{1}{c|}{$J_2$} &
            \multicolumn{1}{c|}{$J_3$} \\
            \hline
            -2.5 & 0.201 & -0.1 & -1.196 & 0.119 & 0.097 \\
            -2.4 & 0.207 & -0.1 & -1.232 & 0.123 & 0.100 \\
            -2.3 & 0.213 & -0.1 & -1.269 & 0.126 & 0.103 \\
            -2.2 & 0.219 & -0.1 & -1.305 & 0.130 & 0.106 \\
            -2.1 & 0.225 & -0.1 & -1.341 & 0.133 & 0.109 \\
            -2.0 & 0.230 & -0.1 & -1.378 & 0.136 & 0.112 \\
            -1.9 & 0.235 & -0.1 & -1.414 & 0.139 & 0.114 \\
            -1.8 & 0.241 & -0.1 & -1.450 & 0.142 & 0.117 \\
            -1.7 & 0.245 & -0.1 & -1.486 & 0.144 & 0.120 \\
            -1.6 & 0.250 & -0.1 & -1.521 & 0.147 & 0.122 \\
            -1.5 & 0.255 & -0.1 & -1.557 & 0.150 & 0.124 \\
            -1.4 & 0.259 & -0.1 & -1.593 & 0.152 & 0.127 \\
            -1.3 & 0.263 & -0.1 & -1.628 & 0.154 & 0.129 \\
            -1.2 & 0.267 & -0.1 & -1.664 & 0.156 & 0.131 \\
            -1.1 & 0.270 & -0.1 & -1.699 & 0.158 & 0.133 \\
            -1.0 & 0.273 & -0.1 & -1.734 & 0.160 & 0.134 \\
            -0.9 & 0.276 & -0.1 & -1.769 & 0.162 & 0.136 \\
            -0.8 & 0.279 & -0.1 & -1.804 & 0.164 & 0.138 \\
            -0.7 & 0.282 & -0.1 & -1.839 & 0.166 & 0.139 \\
            -0.6 & 0.284 & -0.1 & -1.873 & 0.167 & 0.140 \\
            -0.5 & 0.285 & -0.1 & -1.907 & 0.167 & 0.141 \\
            -0.4 & 0.287 & -0.1 & -1.941 & 0.168 & 0.142 \\
            -0.3 & 0.288 & -0.1 & -1.975 & 0.168 & 0.142 \\
            -0.2 & 0.289 & -0.1 & -2.009 & 0.168 & 0.143 \\
            -0.1 & 0.289 & -0.1 & -2.043 & 0.169 & 0.143 \\
            \rowcolor{DMI}-0.0 & 0.289 & -0.1 & -2.076 & 0.169 & 0.143 \\
            & & & & & \\
            \hline
        \end{tabular}
    \end{minipage}
    \label{tab:params}
\end{minipage}
\end{widetext}

\section{Decomposition of the Berry curvature}
\label{sec:berry}

\begin{amendment}
    As we write in the main text, the Hamiltonian of a $\nbands$-sublattice system (here $\nbands = 2$) possesses $2 \nbands$ energies, the first $\nbands$ of which correspond to the particle and the last $\nbands$ to the hole bands.
    The energies of the hole bands at $\vect{k}$ are identical to the energies of the particle bands at $-\vect{k}$ [cf.~Eq.~\eqref{eq:bogtrafo_energies}].
    Although the energies of the hole bands are positive, they are multiplied with the bosonic metric in the expression of the conventional and the generalized Berry curvature:
    $
        \tilde{\varepsilon}_{n \vect{k}}
        =
        \metric_{nn}
        \varepsilon_{n \vect{k}}
    $.
    This is related to the paraunitary normalization of the eigenvectors [cf.~Eq.~\eqref{eq:bogtrafo_paraunitary}], which can also be understood in the framework of Krein spaces~%
    \cite{%
        shindou_topological_2013,
        lein_krein-schrodinger_2019%
    }.
    Because the completeness relation in the Nambu space spans both particle and hole bands,
    $
        \sum_{m = 1}^{2 \nbands}
        \dyad{m \vect{k}}{m \vect{k}}
        =
        1
    $,
    and the velocity operator is in general not (block-)diagonal in the eigenbasis of the Hamiltonian, i.e.,
    $
        (v_{\mu, \vect{k}})_{n, m + \nbands} \neq 0
    $
    for $n, m \leq \nbands$, the Berry curvature of the particle bands includes contributions from the hole bands.
\end{amendment}

\begin{figure*}
    \centering
    \includegraphics[width=\linewidth]{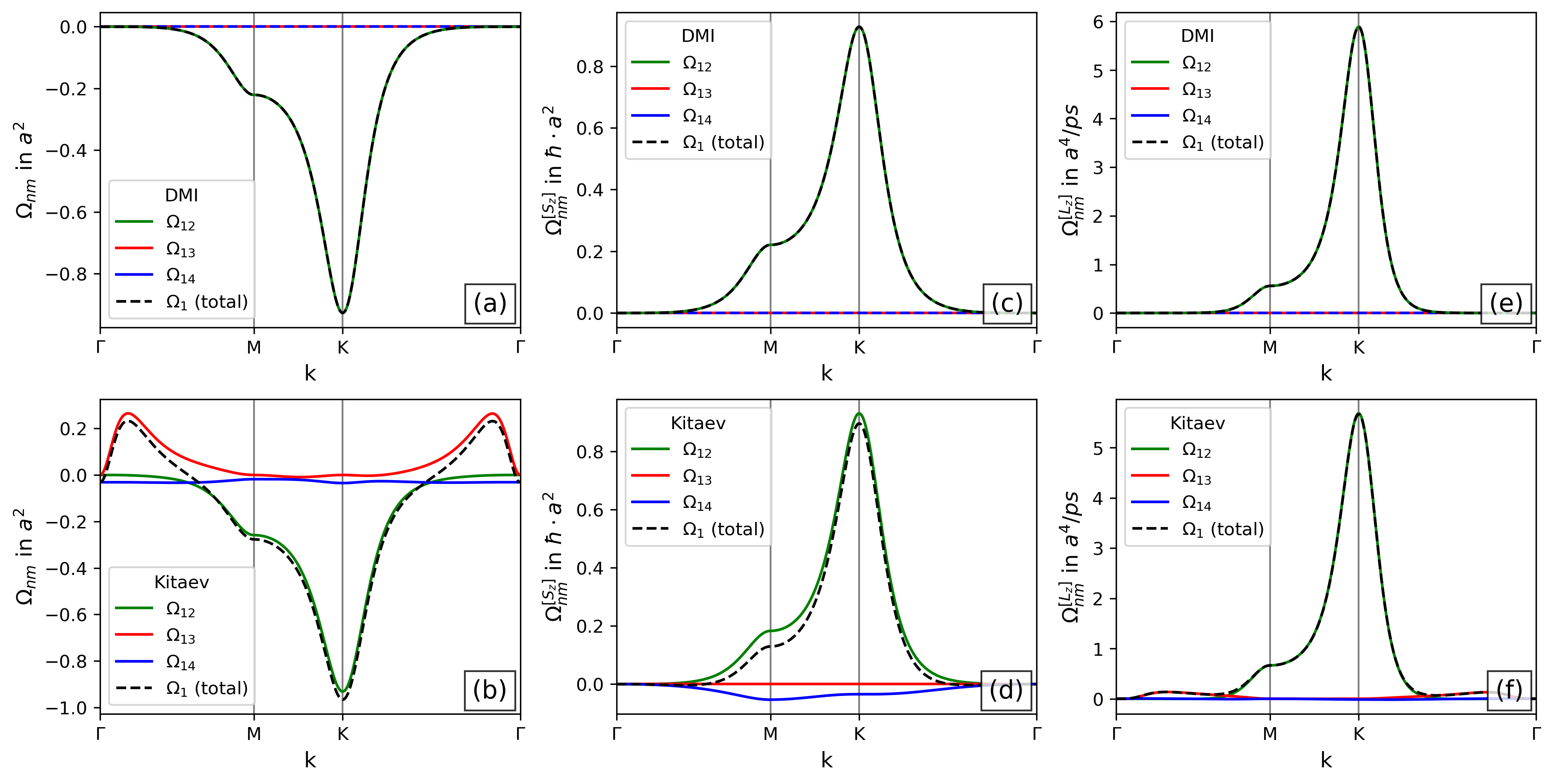}
    \caption{
        Decomposition of the (a,~b) Berry curvature, (c,~d) spin Berry curvature, and (e,~f) orbital Berry curvature of the lowest band $n = 1$ into its individual contributions from the three other bands $m$ for (a,~c,~e) the DMI model and (b,~d,~f) the Kitaev model.
    }
    \label{fig:berry}
\end{figure*}

We have decomposed the Berry curvature $\varOmega_{n \vect{k}}$ into its different contributions $\varOmega_{nm \vect{k}}$ using the definitions in Eq.~\eqref{eq:berry_curv}.
Focusing on the lower band ($n = 1$), the Berry curvature only features contributions from the (particle-like) band $m = 2$ in the DMI model [cf.~Fig.~\ref{fig:berry}(a)].
This is also true for the spin and the orbital Berry curvatures [cf.~Fig.~\ref{fig:berry}(c,~e)].
The reason is that for the DMI model, the magnon number is conserved, which renders the description in Nambu space redundant because the Hamiltonian can be mapped onto a two-band model.

In contrast, for the Kitaev interaction, there are magnon number nonconserving terms that promote additional contributions from the hole-like bands.
For the Berry curvature, in addition to the negative contribution from band 2, there is the positive contribution from the hole-like band 3, which peaks in the vicinity of $\Gamma$ and causes $\varOmega_{1 \vect{k}}$ to reverse [cf.~Fig.~\ref{fig:berry}(b)].
The spin Berry curvature is also mainly dominated by the negative contribution from band 2, but also has a positive contribution from the hole-like band 4 [cf.~Fig.~\ref{fig:berry}(d)].
Lastly, the orbital Berry curvature has two contributions of the same sign from bands 2 and 3 [cf.~Fig.~\ref{fig:berry}(f)].

\section{Spin and orbital currents}
\label{sec:currents}

\begin{figure}
    \centering
    \includegraphics[width=\linewidth]{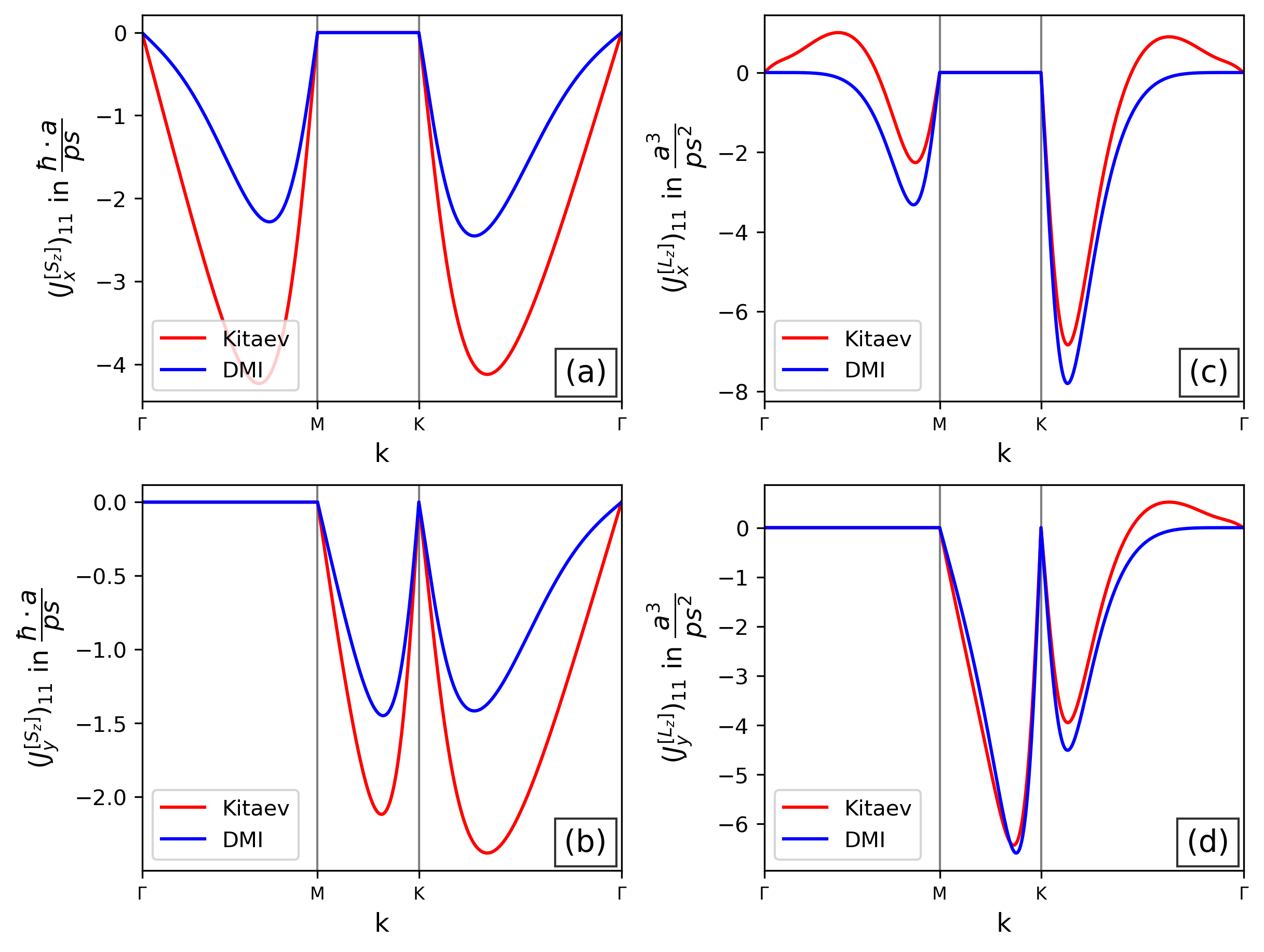}
    \caption{
        Expectation values of the (a,~b) spin current and (c,~d) orbital current along a path in the first Brillouin zone for the lower band.
        Panels~(a,~c) [(b,~d)] display the $x$ [$y$] components of the respective currents.
    }
    \label{fig:current}
\end{figure}

While for the transverse intrinsic transport coefficients the nondiagonal elements of the current operators are essential, the longitudinal transport coefficients only depend on their expectation values (i.e., the diagonal elements).
For the $x$ and $y$ components of the spin current, they are shown in Fig.~\ref{fig:current}(a) and (b), respectively, for the lower band.
The spin current expectation values in the Kitaev model are up to two times larger than in the DMI model, which is caused by the larger spin expectation value brought about by the breaking of spin conservation in the former [cf.~Fig.~\ref{fig:microscopics}(c)].
The expectation values of the $x$ and $y$ components of the orbital current operator, shown in Fig.~\ref{fig:current}(c,~d) for the lower band, do not match the direction of the group velocity in general.
This is because of the complex orbital texture for the Kitaev model that features sign changes [cf.~Fig.~\ref{fig:microscopics}(e)].
For both spin and orbital currents, the $x$ components vanish for both models between M and K because the group velocity perpendicular to the edge of the Brillouin zone is zero.
Their $y$ components vanish between $\Gamma$ and \amend{M since the combined time reversal and two-fold rotational symmetry forces the dispersion relation to be even in $k_y$
[$
    \varepsilon(k_x, k_y) = \varepsilon(k_x, -k_y)
$].}
Thus, the group velocity must be zero for $k_y = 0$.


\bibliography{literature}

\end{document}